\pdfoutput=1
\documentclass[12pt]{article}

\usepackage{pdfpages}
\usepackage{geometry}
\usepackage{subfig}
\usepackage{graphicx}
\usepackage{xcolor}
\usepackage{amsmath}
\usepackage{amssymb}
\usepackage{authblk}
\usepackage[numbers]{natbib} 
\usepackage{subfiles}
\usepackage{dcolumn}
\usepackage{bm}
\usepackage[numbers]{natbib} 
\usepackage{graphicx}
\usepackage{dcolumn}
\usepackage{bm}
\title{ Emergent reprogrammable mechanical memory in soft rods network via friction tuning}
\author[1]{Harsh Jain\thanks{Corresponding author: \texttt{harshjainldh@gmail.com}}} 
\affil[1]{%
Department of Physics, Indian Institute of Science Education and Research, Mohali 140306, India}

\author[2]{ Shankar Ghosh}
\affil[2]{%
Department of Condensed Matter Physics and Materials Science\\ Tata Institute of Fundamental Research, Mumbai 400005, India}

\begin{document}

\maketitle
\begin{abstract}
We present emergent mechanical memory storage behavior in soft cellular materials. The cellular materials are a network of soft hyperelastic rods which store shape changes, specifically local indentation. This happens under an applied global compressive strain on the material. The material transits under strain from an elastic state (capable of 'forgetting' any applied indentation after un-indentation) to a plastic state (indefinitely storing the shape change due to indentation). The memory can be erased via removal of applied global strains and is therefore re-programmable. We characterise this behaviour experimentally and present a simple model that makes use of friction for understanding this behavior. 
\end{abstract}

\section{Introduction}

A child writing on sand at the beach is effectively encoding memory of deformation on the pile of sand. Similarly, a thin long thread placed on a table can be arranged into any shape of our choice. Friction between the neighboring sand particles or between the thread and the table plays an important role here.  The organised state of such a system becomes frictionally locked and therefore capable of storing an encoded ``state", i.e, configuration of constituents until the system is disturbed again by external forces.  A large number of states are accessible to these systems that are meta-stable, i.e., stuck in a local minima. The potential energy landscape for this system is
`rugged'. Such rugged potential energy landscapes are seen in nature by a variety of systems such as jammed systems, glasses and bio-polymer configurations. \cite{samarakoon2016aging, bhatia2022heterogeneity}.
The states of such systems are, however, not protected against
fluctuations or external vibrations and hence the common perception is that these are not useful forms of memory. For any useful form of memory, the essential
characteristics include fast encoding, reading, erasure and long term storage, repeatability as well as low energy dissipation. 
Various physical systems capable of mechanical memory storage are a subject of active interest in the context of reprogrammable braille displays, stably altering mechanical properties of a material etc. \cite{keim2019memory,chung2018reprogrammable,chen2021reprogrammable, winkler2020unveiling,anderson2018glass}. 
Analog Memory storage systems and computation devices for high precision computing, architectures inspired by the physics of neurons in the brain, architectures capable of simultaneous logical computation and memory storage are also gaining interest \cite{markovic2020physics, kingra2020slim, tsividis2018not} along with explorations of emergent functionality in soft robotics and biology\cite{mengaldo2022concise, yang2018synaptic}.
In this paper, we present buckling of a network of rods that are frictionally coupled with the environment and are capable of storing and erasing mechanical deformation locally, subject to global constraints. This permanent
deformation provides a useful model for mechanical memory storage in a
soft medium.
This memory system exploits geometric and mechanical properties of a class of polymeric materials called `hyper-elastic' that show reversible non-linear elastic buckling behaviour to large strains (Compared to metals where permanent deformations occur for strains as low as 0.01). It is lattice-free and analog, i.e., continuously accessible states. It is non-volatile, i.e., no active power required for storage of memory. It is also reprogrammable and allows for encoding and erasure of multiple blocks of memory at once. For long term storage, these polymeric system can be stored in a glassy state below their glass transition temperature. 

\begin{figure}[tb]
	\centering
	\includegraphics[width=0.9\linewidth]{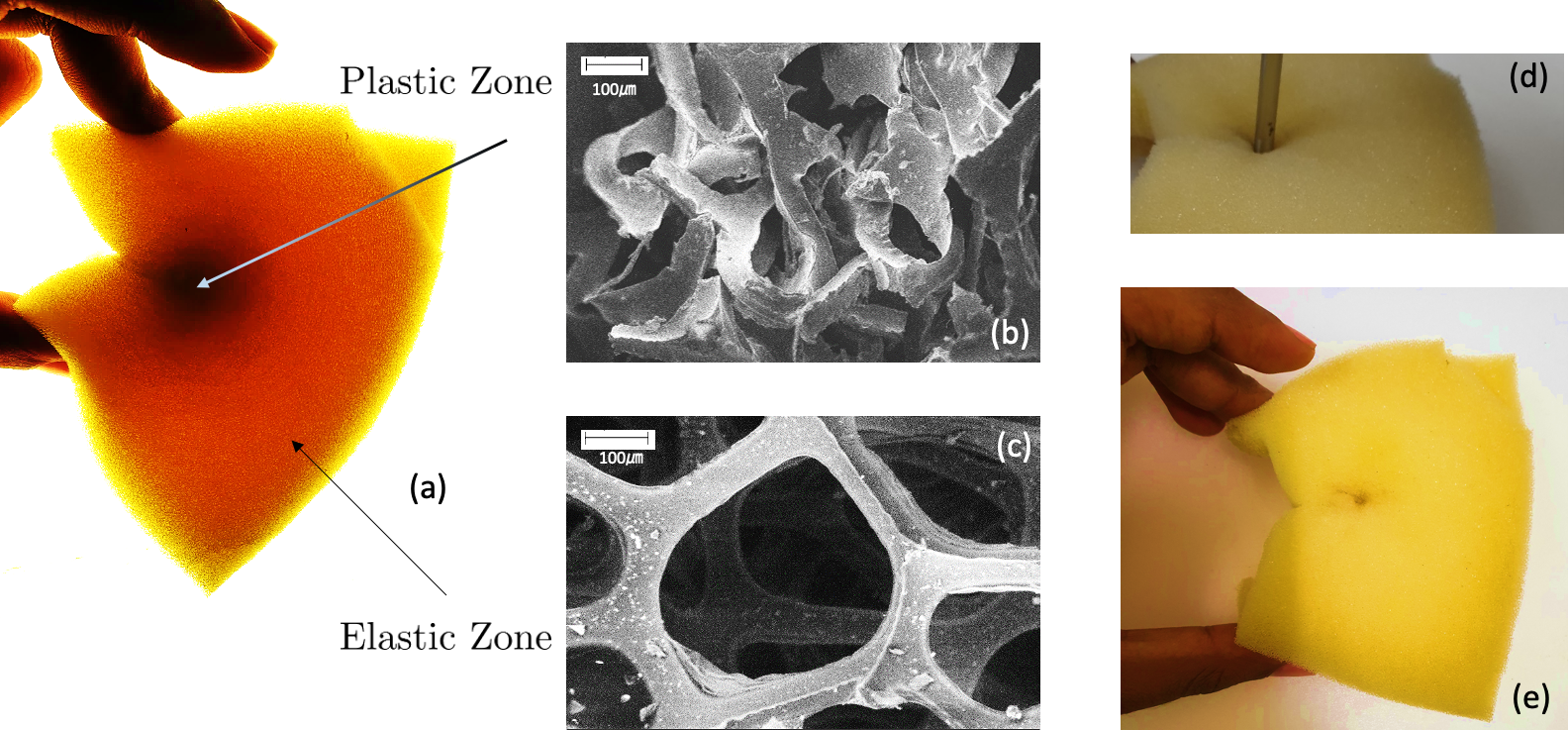}
	\caption[Plastic and elastic zones in sponge, seen in SEM and
        camera images]{(a) shows a cuboidal sponge piece which has been
          compressed at its ends leading to formation of plastic and
          elastic zones as seen with help of background lighting. (b) and (c) show the SEM (Scanning electron Microscope) images of the plastic and elastic regions respectively. (d) shows the mechanism of indentation with a metallic rod and (e) shows the persistence of memory after removal of indentor due to frictional coupling in the plastic region}
	\label{fig:proposal}
\end{figure}

Figure \ref{fig:proposal}(a) depicts a familiar sponge found in our daily lives, comprising soft polymeric rods. These rods are interconnected in three-dimensional space, forming a random network known as a cellular network, as described in literature \cite{gibson2003cellular}. 
The polyurethane sponges utilized in our experiments possess low elastic moduli,  $E\sim 1 \text{MPa}$ , enabling easy compression between two fingers. Consequently, this compression leads to the creation of a distinct region within the material known as the ``Plastic Zone," as observed by the darker region in Figure \ref{fig:proposal}(a). The remaining portion of the material is referred to as the ``Elastic Zone." To visually demonstrate these regions, we backlit the material using white light. Detailed scanning electron microscope (SEM) images of these zones are displayed in Figure \ref{fig:proposal}(b) and (c).
In the Elastic Zone, the interaction between each rod and the rest of the network primarily occurs at it's end points through mechanical joints, as illustrated in Figure \ref{fig:proposal}(c). In the Plastic Zone, each rod adopts a buckled configuration leading to dense packing. Figure \ref{fig:proposal}(b) demonstrates that in this zone, each rod establishes surface contact with multiple neighboring rods in the network.

Indenting with a metallic indentor, gives quite distinct response in both of these zones. In the elastic zone, the sponge regains its original shape following unindentation, leaving no trace of deformation. In contrast, the plastic zone retains a permanent deformation depending on the depth of indentation. However, this deformation memory can be erased by allowing the plastic zone to relax back into the elastic zone through the removal of applied global stress on the sponge.

In the following section, we characterize our mechanical memory storage system experimentally. 
\section{Experimental Details}
 
\begin{figure*}[t]
	\centering
	\includegraphics[width=0.7\linewidth]{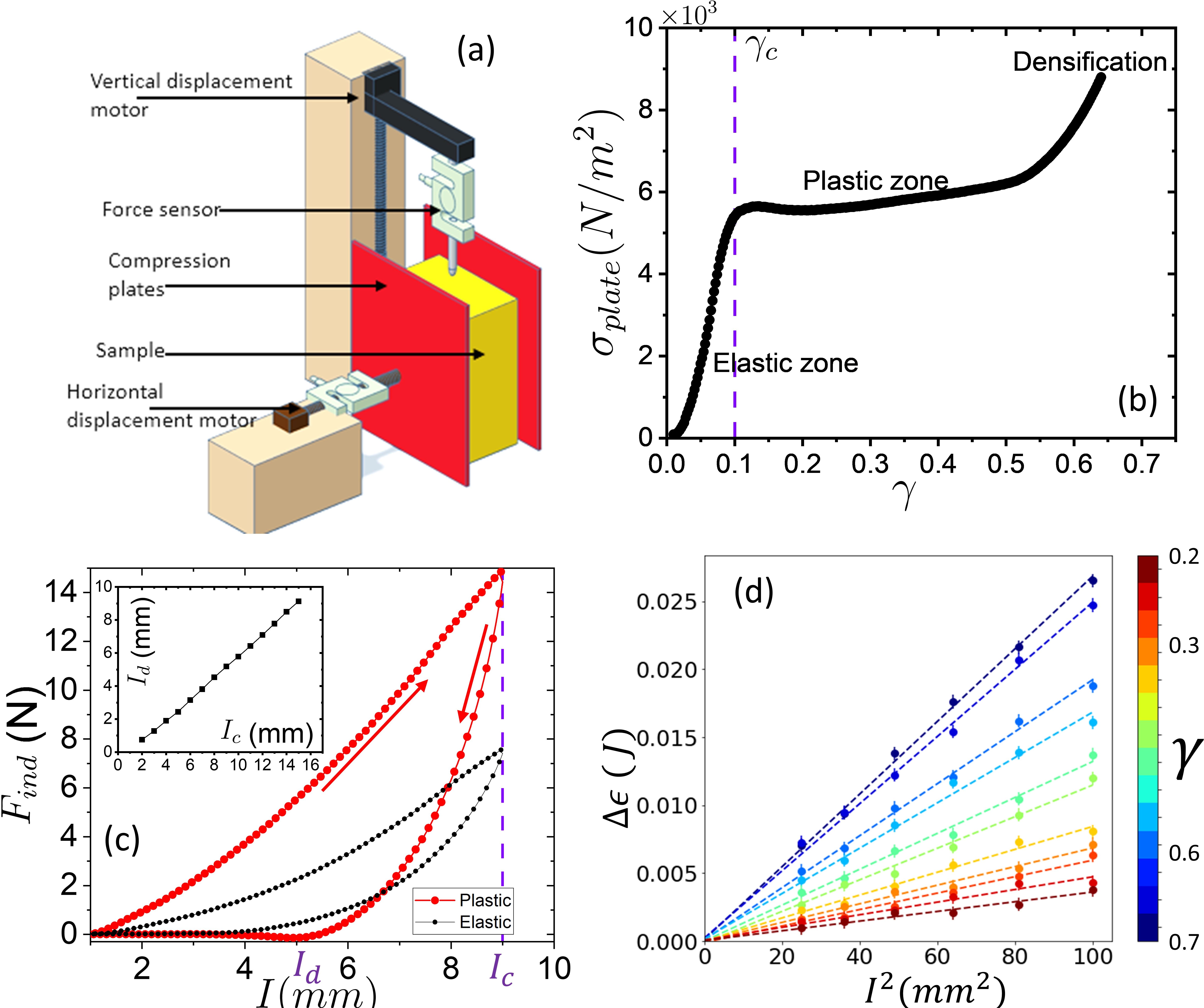}
	\caption[Experimental details for the compression of sponge
        network] {(a) shows a 3D schematic for the experimental setup that we designed, consisting of motorized horizontal compression and vertical indentation via an indentor coupled to motor via a force sensor.  (b) shows the $\sigma_{plate}$ vs $\gamma$ curve for compression of sponge with $E\sim 7MPa$ between plates with the critical strain $\gamma_c$ shown with the violet vertical line. (c) shows the hysteresis curve ($F_{ind} vs I$) for indentation in the elastic region (black) and plastic region (red). Arrows show the forward journey (indentation) and the return journey (unindentation). $I_d$ and $I_c$ for the plastic cycle are shown on the x-axis. Inset of (c) shows the variation of $I_d$ vs $I_c$ for the plastic zone. (d) shows the variation of hysteresis ($\Delta \epsilon$) with $I^2$ for various strain values ($\gamma$) represented by the color-map. }
	\label{fig:ExptSetup}
\end{figure*}

Figure \ref{fig:ExptSetup}(a) illustrates the experimental setup employed to subject the sponge samples to strain $\gamma$ and 
 depth of indentation $I$. To induce varying levels of plasticity in the sponge, we horizontally compressed a  $ 10 \textrm{cm} \times 10 \textrm{cm} \times 10 \textrm{cm}$ cube of sponge  between two metallic plates. One of the metallic plates was mounted on a motorized stage, allowing horizontal movement. For the indentation process, a metallic rod with a diameter of approximately $5 \textrm{mm}$ was used as the indenter. The indenter  was  mounted on a motorized stage. This configuration provided precise control over the  vertical movement of the indenter into and out of the sample.  Force sensors were attached to both the moving horizontal plate and the indentor. Throughout the  compression and decompression process of the sponge and the loading and unloading  procedure of the indenter, the forces were measured using two Kiethley 2010 multimeters. In our experiments, we defined the indentation depth as zero when the indenter initially made contact with the unindented sponge.

The stress-strain response of the sample during compression between the horizontal plates is depicted in the inset of Figure \ref{fig:ExptSetup}(b). The experiment begins  with the plates separated by  $L \approx 10 \textrm{cm}$. When subjected to low levels of compressive strain, $\gamma = \Delta L/L$, where $\Delta L$ represents the extent of compression, the material demonstrated elastic and linear behavior. 

Once the strain surpasses a critical value of approximately $\gamma_c \sim 0.1$, specific regions within the sponge experience collapse, leading to the formation of a localized plastic zone. With further increase in strain, this plastic zone expands in size.  The signature of the formation of this plastic zone is borne in the  measured stress-strain curve  where  beyond $\gamma_c$   the strain continues to rise without a significant increase in stress. The sponge now enters a uniformly plastic zone. Following this, a densification process begins where the compression results in denser packing of the rods. In this phase of densification the stress begins to  again increase with the strain. This behavior is consistent with observations in a broad range of cellular materials \cite{gibson2003cellular}.

The experimental samples consisted of hyper-elastic cellular sponges made of polyurethane rods, exhibiting elasticities ranging from $1 \textrm{MPa}$ to $10 \textrm{MPa}$.Individual rods are polydisperse with average radius  $r\sim 0.05mm$ and length $\ell \sim 0.5mm$. The polyurethane material displays viscoelastic behavior, wherein the viscous response due to indentation gradually relaxes over a characteristic time scale (see Suppl. Information). To specifically investigate the elastic response of the material,  we performed slow  indentation. In our experiments, this indentation rate was limited to approximately $0.1 \textrm{mm/s}$.

We employed two protocols to assess the mechanical response of the sponge under indentation deformation. The  protocol $\mathbf{P}_1$ involves erasing all previous system memory before conducting each indentation experiment. This is achieved by resetting the compressive strain $\gamma$ to zero after the completion of each experiment. Subsequently, this pristine system is compressed to the desired $\gamma$ value, and the indentation experiments are carried out.

The  protocol $\mathbf{P}_2$, reminiscent of measuring minor hysteresis loops in magnets, begin with compressing the material to a predetermined strain level from its pristine state ($\gamma=0$). Next, the material is indented to a small indentation value $I_c$ and subsequently un-indented. As we progressed through subsequent indentation loops, the value of $I_c^n$ was gradually increased.   We define the maximum indentation depth in a given indentation cycle as $I_c^n$ for the n\textsuperscript{th} cycle, here $I_c^n<I_c^{n+1}$. Throughout the entire process, we consistently measured the indentation depth and the corresponding force of indentation $F_{ind}$. Importantly, the compressive strain was maintained at the desired value throughout the entirety of the process.

Figure \ref{fig:ExptSetup} (c) presents a graph depicting the relationship between the applied force by the indentor $F_{ind}$ and the depth of indentation $I$ under protocol $\mathbf{P}_1$ . This graph showcases the occurrence of hysteresis, which is a characteristic of the viscoelastic response of the material. The black curve represents the response at a low strain value of approximately $\gamma \approx 0.1$ The sponge here is  in a  predominantly elastic state. The hysteresis observed in this region is attributed to the viscoelastic behavior of the media.
On the other hand, the red curve corresponds to the response at a high strain value of approximately $\gamma \approx 0.7$. The sponge is in its plastic state. The hysteresis observed here is associated to the frictional interaction between neighboring collapsed rods within the material. Arrows have been used within the figure to provide visual cues  for the indentation and un-indentation processes. During the return cycle marked by the arrow pointing right to left in Fig.\ref{fig:ExptSetup} (c), we  observe that there  exists a residual indentation $I_d$ in the system.  It is important to note that unless this residual indentation memory is eradicated, it exerts a significant influence on subsequent indentation cycles, as we will explore later in the paper.

The region between the forward path and the reverse path on the $F_{ind}$ versus $I$ curve represents an area denoted as $\Delta \epsilon$, which is measured in units of energy (Joules). This area quantifies the energy dissipated during the process. Fig. \ref{fig:ExptSetup} (d) demonstrates that the magnitude of $\Delta \epsilon$ increases quadratically  as the depth of indentation $I$ increases.

\begin{figure*}[t]
	\centering
	\includegraphics[width=0.99\linewidth]{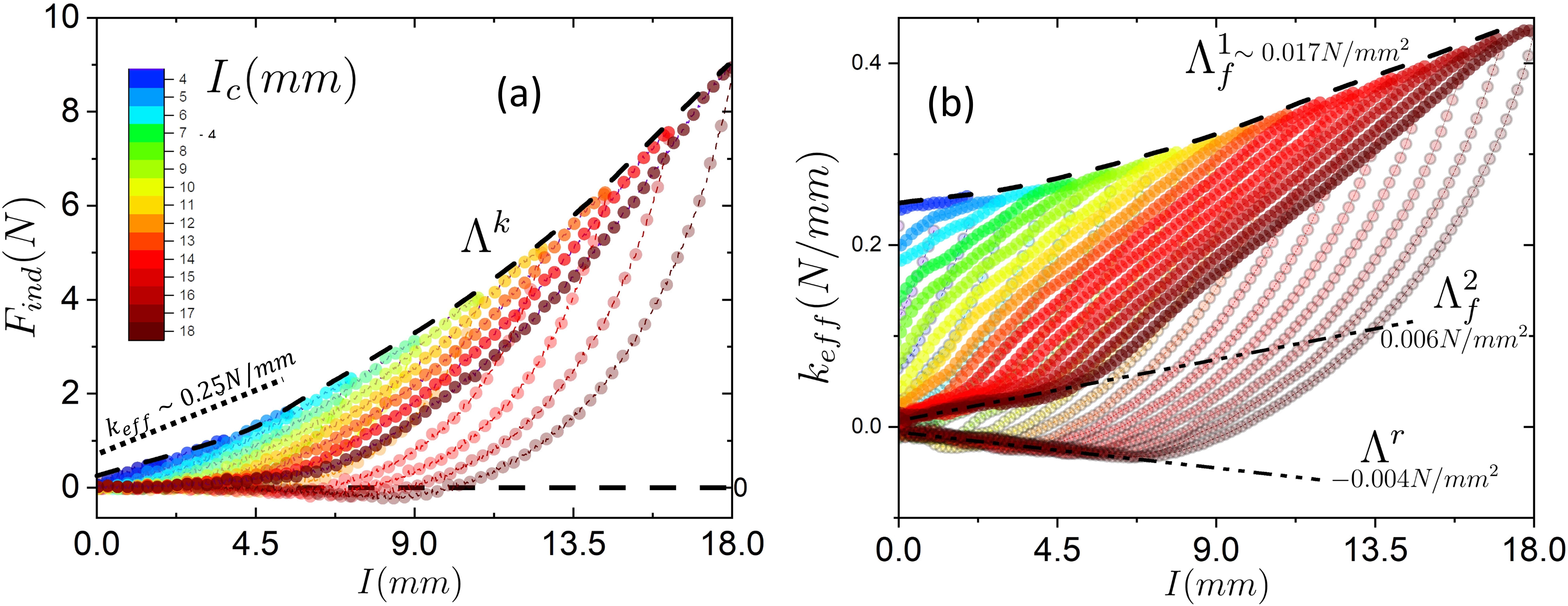}
	\caption[$F_{ind}$ v/s $I$ depth for compressing rod network] {(a) For a sponge with elasticity $E\sim 7 MPa$, the forward paths and return paths are shown on the $F_{ind}$ vs $I$ curve by the darker shade and lighter shade filled circles respectively. Corresponding maximum indentation values ($I_c$) are shown on the colormap. (b) shows the $k_{eff}$ vs $I$ plot for the same data. Experiments are performed with protocol $\mathbf{P}_2$ and $\gamma\sim 0.7$. Envelopes $\Lambda_k$ and $\Lambda_f^1$ show the data for experiment with protocol $\mathbf{P}_1$ as discussed in the text; Envelopes $\Lambda_f^2$ and $\Lambda_r$ appear as a result of frictional force experienced by the indentor in the forward and return journey respectively. }
	\label{fig:forwardbackwardcycle}
\end{figure*}

Figure \ref{fig:forwardbackwardcycle} displays the outcomes of multiple indentation cycles performed following protocol $\mathbf{P}_2$. Throughout these cycles, the maximum indentation depth $I_c$ is systematically varied from $1\, \textrm{mm}$ to $18\, \textrm{mm}$ in increments of $1\, \textrm{mm}$, while maintaining a constant compressive strain $\gamma$ of 0.7. Figure \ref{fig:forwardbackwardcycle} (a) illustrates the $F_{ind}$ versus $I$ paths for both the indentation and un-indentation processes, with a focus on specific representative values of $I_c^n$. Additionally, in Figure \ref{fig:forwardbackwardcycle} (b), we present the corresponding instantaneous effective  material spring constant $k_{eff} = F_{ind}/I$ of the system.
The black dashed lines form an envelope depicting the trajectory that the $F_{ind}$ versus $I$ curve would have followed  during the loading process if no previous deformation had been stored.  The  memory of the past deformation, lowers the value of the force experienced by the intendor. It is only on  surpassing the previous cycle's $I_c^{n-1}$ that the $F_{ind}$ versus $I$ curve retraces the path along the above mentioned black envelope for the subsequent deformation.  
In contrast to an elastic spring where the force increases linearly with displacement, the black dashed line marked as $\Lambda^k$ in Fig. \ref{fig:forwardbackwardcycle}(a) serves as the envelope in this case exhibits a superlinear growth pattern as the indentation progresses. This behavior is  also evident in the fact that  the upper envelope  $\Lambda_1^f$ of $k_{eff}$, as depicted in 
Figure \ref{fig:forwardbackwardcycle} (b) demonstrates an increase in response with indentation. Notably, two distinct envelopes of $k_{eff}$ can be observed. The lower envelope, denoted as $\Lambda_2^f$, corresponds to lower indentation values, while the upper envelope, referred to as $\Lambda_1^f$, corresponds to higher indentation values. 

The lower envelope, denoted as $\Lambda_2^f$, emerges as a result of the frictional interactions encountered by the indenter while passing through the depression, reminiscent of the previous indentation cycle. Conversely, the upper envelope, $\Lambda_1^f$, becomes apparent at higher indentation values, where the indenter ploughs through the undeformed region of the sponge. In this process, the indenter experiences mechanical resistance caused by the deformation of the sponge.

During the return cycles, as the indenter undergoes unloading, the force of indentation ($F_{ind}$) gradually diminishes until it reaches zero at a specific indentation value $I_d$. It is this deformation corresponding to $I_d$ that is retained as long-term memory. As the loading process begins, the sponge grips the surface of the indenter, generating a frictional force that must be overcome during unloading. This frictional force manifests as a negative force acting on the indenter. Importantly, the strength of this frictional grip intensifies with increasing compressive strain ($\gamma$).

For low indentation values, all the return traces of the indentation force lie on the same curve. This behavior is reflected in the asymptotic envelope denoted as $\Lambda^r$ observed for the return curves. Along this trace, the effective stiffness ($k_{eff}$) is negative, indicating the presence of a frictional force acting opposite to the direction of the indenter's velocity. Notably, the force of friction amplifies as the indentation extent increases.

Both $\Lambda_2^f$ and $\Lambda^r$ exhibit similar slopes since they originate from frictional interactions between the indenter and the sponge.

It is important to highlight that the curves depicted in this study exhibit a high level of repeatability, indicating consistent results across multiple indentation cycles. Furthermore, there is no observable aging of the sponge material as we progress through these cycles. Supplementary data illustrating the fluctuations in $F_{ind}$ for multiple runs is provided in the SI, revealing distinct signatures of the microstructural arrangement of the network that remain consistent throughout. This substantiates the absence of any noticeable aging effects within the system.

\begin{figure*}[tbp]
	\centering
	\includegraphics[width=0.9\linewidth]{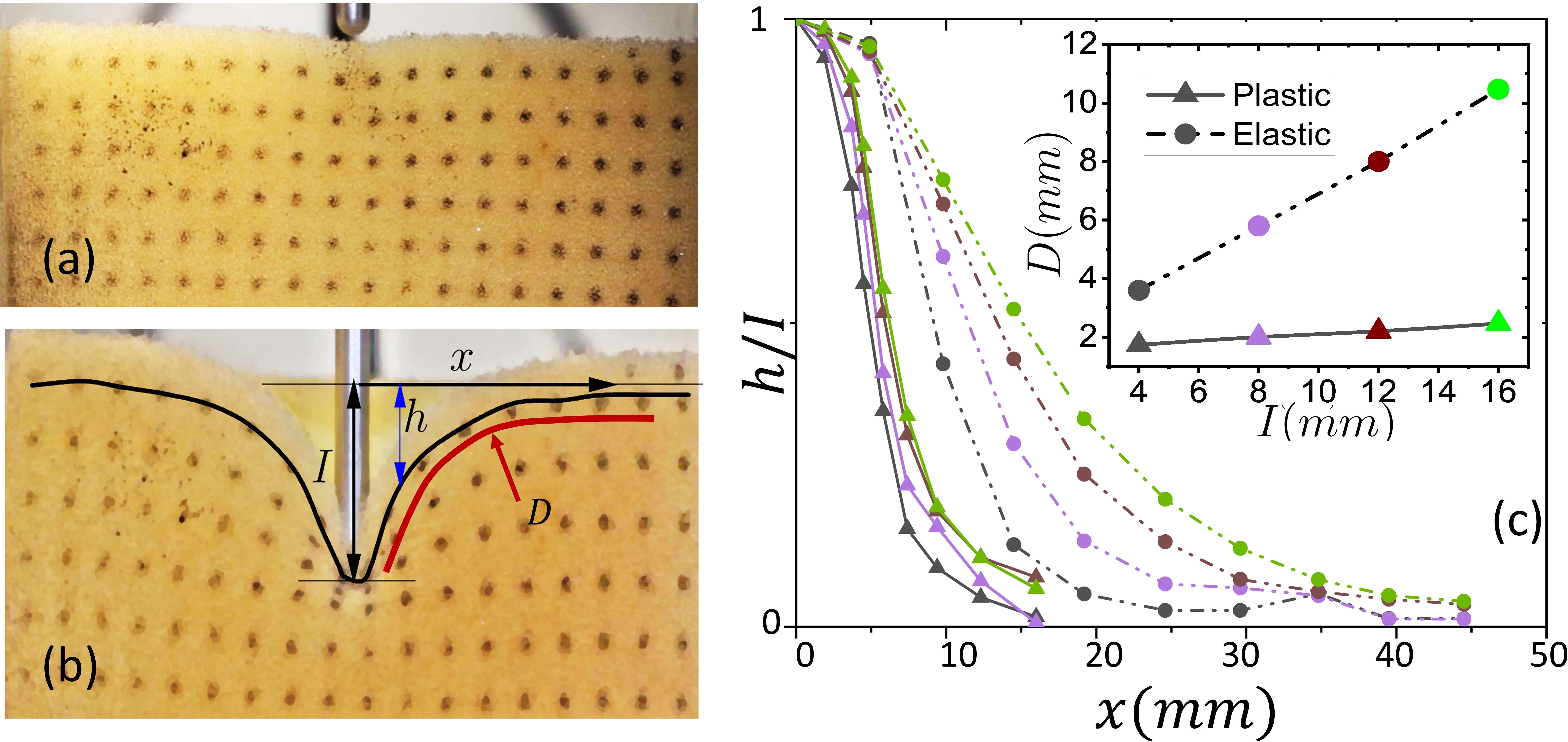}
	\caption[localisation] {Figure (a) shows a side-view of the undeformed sponge with a regular grid pattern. (b) shows the side-view of spatial decay of deformed height $h$ as a function of distance from indentor $x$ fitted by an exponential with parameter $D$ such that $h(x)=I exp(-x/D)$ where $I$ is the current value of indentation by the metallic indentor. (c) shows the various profiles $h/I$ in the plastic as well as elastic state for various indentation values. The indentation values $I$ and the corresponding fitted parameter $D$ are shown in the inset. }
	\label{fig:local}
\end{figure*}

Figure \ref{fig:local} depicts images of the sponge sample during loading and unloading by the indenter, showcasing both the elastic and plastic zones. To facilitate visualization in the cross-section geometry, the indentation is conducted near the edge of the sponge sample. A grid measuring $5 \, \textrm{mm} \times  5 \, \textrm{mm}$ is imprinted on the sponge to aid in tracking the strain field as shown in Fig. \ref{fig:local}(a). The case of elastic loading is shown in (b) with $h, x, I$ and $D$ labelled such that $h(x)=I \exp(-x/D)$. The profiles of $h/I$ vs $x$ in the plastic and elastic states are shown in Fig. \ref{fig:local}(c) demonstrating that the deformation resulting from indentation is more localized  in the plastic zone as compared to the elastic zone. The fitted values of $D$ and the corresponding $I$ values are shown in the inset.

\section{Model}
\begin{figure*}[t]
	\centering
	\includegraphics[width=0.7\linewidth]{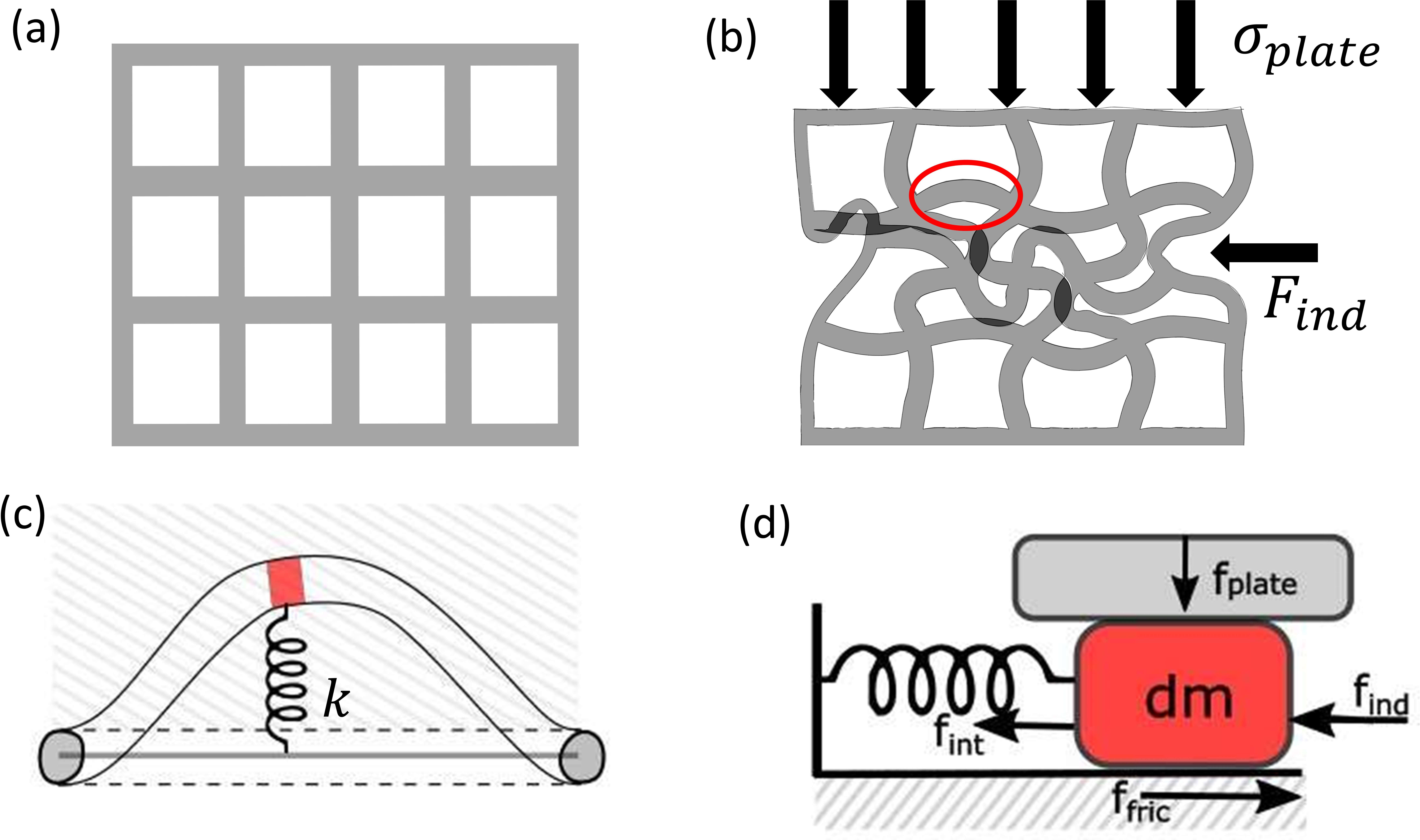}
	\caption[Drawing of configuration of the rods]{(a) shows an
		undeformed configuration of rods network in 2D (b) shows the deformed
		configuration of a rod similar to configurations in Fig. 1 obtained via simulations for hyperelastic material (c) shows the undeformed (dotted) and the deformed (solid) configuration of a single rod and a small element $dm$ (red) of the rod under restoring spring force (d) shows a diagram of various small forces on the small element $dm$}
		\label{fig:model}
\end{figure*}

 Fig. \ref{fig:model}(a) shows an approximation for a two-dimensional regular sponge, made up of a network of 
hyperelastic rods. 
Fig. \ref{fig:model}(b) shows the deformed configuration under applied stress $\sigma_{plate}$ obtained via simulations for hyperelastic material as detailed in the Supplementary Information. A small region of this deformed configuration circled in red in (b) is shown in Fig. \ref{fig:model}(c). The unbuckled configuration has been shown in a dashed line and the corresponding buckled configuration of the rod is in solid black line. The restoring forces due to elastic deformations can be captured with an effective spring constant $k$ \cite{jain2021compression}. Each small section of the rod shown in Fig. \ref{fig:model} (c) with mass $dm$ (shown in red) is in  mechanical equilibrium under various forces. Highlighted in Fig. \ref{fig:model}(d) are the frictional force ($f_{fric}$), force by the compressing plate ($f_{plate}$), force due to indentation ($f_{ind}$) and internal elastic restoring force ($f_{int}$). The weight of the material is balanced by the internal forces and hasn't been shown here explicitly. Here, $f_{int} = kx$ where $k$ is the spring constant and $x$ is the displacement from equilibrium position $x=0$. 
The frictional force $f_{fric}$ = $\mu f_{plate}$ is a result of applied compressive force $f_{plate}$ which captures the behaviour of $\sigma_{plate}$ in Fig. \ref{fig:ExptSetup}(b), and $\mu$ is the coefficient of the frictional interaction between adjacent rods. From our observations, the value of applied stress $\sigma_{plate}\sim 8000 N/m^2$ from Fig. 2(b) is found to be close to the value of frictional stress experienced by the indentor $\Lambda_r \sim 4000 N/m^2$ which gives us an estimate of $\mu \sim \Lambda_r/\sigma_{plate} \sim 0.5$.

The condition $|f_{fric}|=0$ characterizes the elastic state as there are no neighbouring contacts. While indenting in elastic zone with an indentation force $f_{ind}$, the block gets displaced from it's equilibrium position $x=0$ to a new equilibrium position is $x_c$ . Upon removal of $f_{ind}$, the block returns to its equilibrium position at $x=0$ from $x=x_c$. Consequently, in this case, the system retains no memory of the indentation. 
The plastic state is described by a finite friction force $|f_{fric}|=\mu |f_{plate}| > 0$.   
The equilibrium of the forces on the block is given by $f_{ind}= f_{fric} - k x_c$, i.e., the equilibrium position is 
\begin{equation}
x_c = 
	-\frac{f_{ind}- \mu|f_{plate}|}{k}.     
 \label{Eqn:1}
\end{equation}

On removing the indenting force $f_{ind}$,  the block retraces its path as long as the inequality,  $f_{ind} > \mu(x) f_{plate} +  kx$, is satisfied. The block stops at 
\begin{equation}
x_d = \begin{cases}
	x_c &  |k(x_c)|<\mu |f_{plate}|\\
	-f_{fric}/k & |k(x_c)|>\mu |f_{plate}|
\end{cases}
\label{Eqn:2}
\end{equation}

Therefore, $x_d$ and $x_c$ essentially capture the behaviour of $I_d$ and $I_c$ from our experiments. 


\section{Conclusions}
We find the soft hyperelastic rods be an appropriate system for displaying emergent features such as storing the memory of past deformations. Low bending stiffness and high frictional interaction as well as hyperelasticity are necessary from the proposed memory model. High friction between hyperelastic materials is largely unexplored and recent studies have tried to probe the phenomena \cite{liang2023friction}. The features discussed have little dependence on the scale of the material or it's constituents and more on it's geometric and mechanical features.
In more finely controlled systems, more states can be encoded in the orientational microstructures increasing the memory density. Polymeric materials can be designed to have a huge range of glass transition temperature ($T_g$) and the states can be encoded above $T_g$ and stored below. 
Our work provides a physical picture that is applicable at the biological scale as well, given that neuron synaptic structures have elasticity of similar magnitude $\sim 500kPa$ and properties at the synaptic junction are regulated by stiffness control of actomyosin network in the pre and post synapse via  binding proteins\cite{yang2018synaptic}. A wide variety of artificial rods as
well as biological polymers can demonstrate the buckling features we discussed at a variety of scales \cite{peacock2020buckling,xiang2011predicting,shaw1966plastic,dai2016three}. 

\textbf{Acknowledgements:} We would like to thank by Prof. Andy Ruina and Prof.Krishnaswamy Nandakumar for useful discussions and insights.

\bibliographystyle{unsrt}
\bibliography{main}

%
\pagebreak


\section*{Supplementary material}

# Supplementary Information

# Contents

# 1 Further Experimental Data

## 1.1 Features of Memory

Fig. 1 shows images of a sponge of elasticity $E \sim 7MPa$. We show response in the elastic and plastic state in Fig. 1(a) and (b) respectively. (a) shows no deformation $I_d$ is stored whereas, on indentation to $I_c$, as shown in Fig. 1 (b)(ii) and the indentation up to a depth $I_d$ is permanently stored as shown in (b)(iii). Fig. 1(c) shows that small indentation with various indentors can be created on the surface in distinct shapes, which can be thought of as writing multiple bits in parallel and close to each other. Also, as we discussed in the main text that the deformation in the plastic state is highly localised, a demonstration of this can be seen in Fig. 1(d). A small hexagonal unit of a few rods is colored blue as shown in (d) (i). It is then indented by a small indentor of diameter 2mm and the state of the colored unit of rods

after indentation is shown in (d)(ii). In our experiments, we also observed that applying strain by compressing on both orthogonal axes perpendicular to the direction of vertical indentation leads to even better retention of the deformed shape. However, most features discussed are present in the case of uniaxial strain. So, we have focused only on the case of uniaxial strain in our experiments.

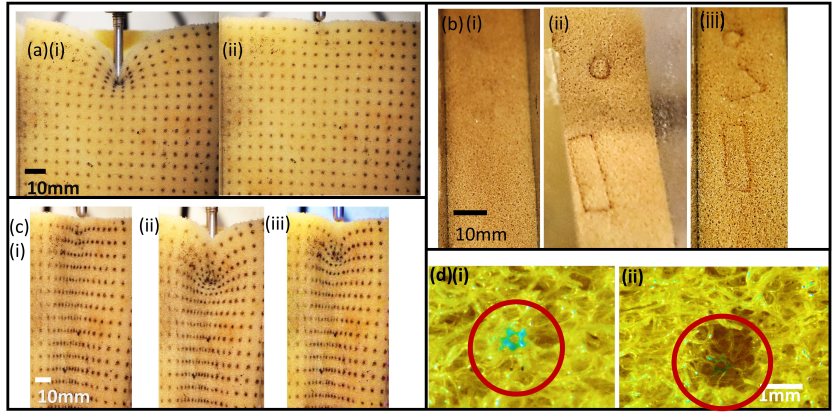

Figure 1: (a)(i) shows the indented elastic state of sponge and (ii) shows that on removal of indenatation, a regular grid printed on the undeformed sponge comes back to it's original state. (b)(i)-(iii) show the sponge in plastic state before, during and after indentation respectively. (c) shows various shapes that can be indented via differently shaped indentors. (d) (i)-(ii) show the behaviour of a small blue colored region (circled red) in the plastic state before and after indentation as seen with a 10x microscope.

## 1.2 Reproducible Fluctuations in $F_{ind}$

We discuss the fluctuations in the measured value of $F_{ind}$ here. It was found that small fluctuations in the value of $F_{ind}$ are repeatable in multiple experiments of indentation, when performed at the same location. This reproducibility is demonstrated in Fig. 2. This can be attributed to re-formation of the complex buckled microstructure of the rods. This therefore substantiates the claims of the complete erasure of previously stored memory on setting $\gamma = 0$.

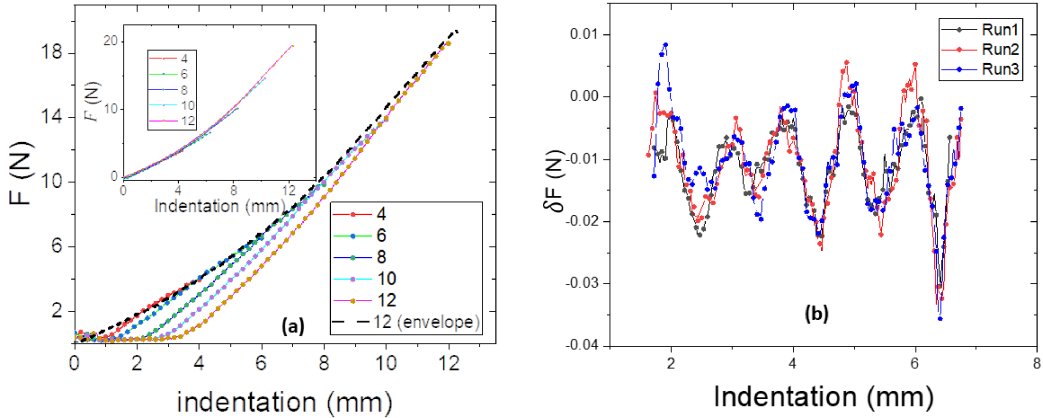

Figure 2: (a) shows only the forward part of indentation cycle for indentation to various depths following protocol $\mathbf{P_2}$. (b) shows $F - F_{avg,T}$ for the indentation force.

## 1.3 Sponge properties and characterization

Cellular materials have a vast literature discussed in [1]. These can be broadly classified as open-cell or close cell foams or sponges. The cellular materials discussed in this paper are of open-cell type, i.e., made of rods, while closed-cell sponges are made of walls rather than rods. These materials demonstrate interesting negative Poisson's ratio as demonstrated in Fig.3 Considering $h(x) = 0$ to be the initial profile of the surface where $x$ varies from 0 to $L$ ($L$ being the distance between the compression plates). It goes to a slightly u-shape with $h(x) < 0$ for finite strains. This is typical behaviour known in literature for cellular materials[1]. The rods at the surface are buckled in random orientations and on average, the surface is smooth.

Since these polymeric materials are viscoelastic, there is a time rate at which applied forces decay as seen in Fig.3(b). We see that if compressed at a high strain rate, $F_{ind}$ will relax in time, with a very short timescale. Whereas, at low enough strain rates, relaxation time will be very long, and force relaxation will be negligible.

# 2 Theoretical Analysis

We now start our discussion with the behaviour of a single hyperelastic rod under buckling.

iii

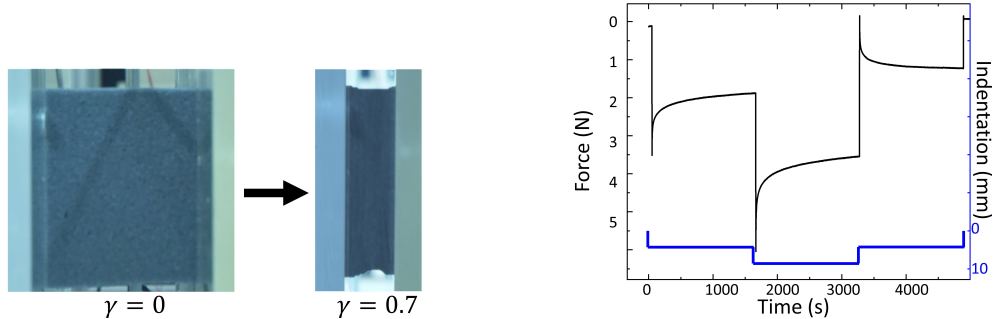

(a) Figure shows that height after compression decreases, implying negative Poisson's ratio.

(b) Figure shows the variation of response of sponge in the plastic state under fast indentation with time.

Figure 3: Sponge Properties

## 2.1 Buckling of a single rod versus network

A single rod will undergo Euler buckling at critical load $n^2\pi^2 EI/L^2$

Fig. 4 (a) shows the undeformed configuration of a 3D rod parameterised by arclength $s$ ($s$ varies from 0 to $\ell$) along the center of the rod on the 'centerline' in Euclidean $R^3$ space. At every point $s$ on the rod, we can associate three orthonormal vectors $\vec{d_1}, \vec{d_2}$ and $\vec{d_3}$ which stand for Tangent, Normal and Binormal. Together these are also known as the Serret-Frenet triad. In the deformed condition of the rod shown in Fig. 4 (b), the vectors will continue to be orthonormal if we ignore the shearing of the rod (which holds for for thin rods). For uniform deformation, we can account for the stretching, bending and twisting of these rods via the derivatives of the vectors $\vec{d_1}, \vec{d_2}$ and $\vec{d_3}$ along the parameter $s$. For orthonormal vectors, the derivatives can be broken down to three scalar functions $\kappa^{(1)}(s), \kappa^{(2)}(s)$ and $\tau(s)$ [2]. Here $\kappa^{(1)}, \kappa^{(2)}$ and $\tau$ are interpreted as the local rates of rotation of the Serret-Frenet triad. $\kappa^{(1)}$ and $\kappa^{(2)}$ represent curvature due to bending of the rod along the plane orthonormal to the tangent vector, and *tau* refers to the twisting of the rod, around the tangent vector. Also, here $\int ds = \ell + d\ell$, and $u = d\ell/\ell$ is the uniform strain in the rod. The energy per unit length for the rod in terms of curvature, twist and strain can be written as

$$\frac{\epsilon_{rod}}{\ell} = \int ds \frac{EI}{2}\kappa^{(1)}(s)^2 + \int ds \frac{EI}{2}\kappa^{(2)}(s)^2 + \int ds \frac{\beta J}{2}\tau(s)^2 + \frac{E\pi r^2}{2}u^2 \quad (1)$$

Here, $I = \pi r^4/4$ and $J = \frac{\pi r^4}{2}$ are moments of bending and twist and $\beta$

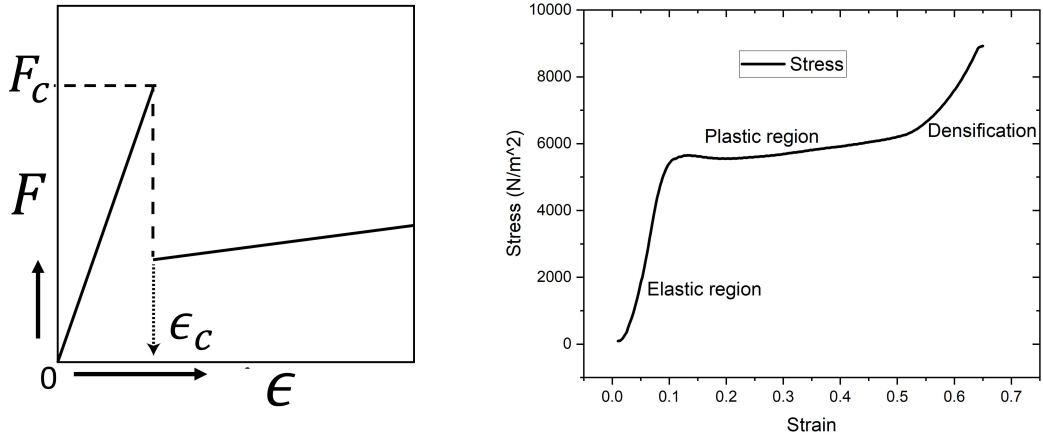

Figure 4: (a) shows the $F$ v/s $\epsilon$ curve expected for buckling of a single rod. (b) shows the stress-strain curve for a network of rods as observed in experiments.

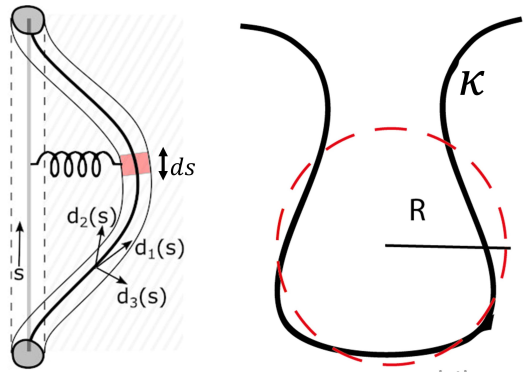

Figure 5: (a) shows a theoretical model of buckling (b) shows the order of magnitude of maximum curvature in the rod..

is the mass per unit length. We find that elastic energy and bending energy scale as $r^4$ whereas stretching energy scales as $r^2$. This implies that for thin rods, stretching is much more costly, and the stretching restoration forces are much higher than bending restoration forces. Resistance to bending can be thought of as being provided by a spring constant $k_t$ and resistance to stretching is provided by a spring constant $k_\ell$. The buckling and the stretching spring constants are related by $k_t/k_\ell = (r/\ell)^2$ which points to an arguement for inextensibility in the rod and a treatment of an individual rod similar to elastica. Another argument for the same can be seen below.

## 2.2 Inextensibility condition

Let $F$ be the external load applied to a rod. For moderate loading $F$, where forces are close to the weight of the rods, the buckled rods won't contain sharp kinks and the curvatures are shown in Fig. 4 (d). The buckling moment is $FL \sim EI\kappa$. Any strains under an axial load $F$ will be given as $\epsilon \sim F/(AE)$ where $A = \pi r^2$ is the cross section. Eliminating $F$, we observe that

$$\frac{\epsilon}{\kappa} \sim \frac{r^2}{L} \qquad (2)$$

From Fig. 4 (d), we can estimate that the typical curvature $\kappa \sim 1/L$. Therefore, $\epsilon \sim (r/L)^2$.

Hence, for slender rods, the inextensibility condition holds. This is true when any indentor force is removed and all the forces in the system are internal forces.

Under a large force, sharp kinks can be obtained in the curvature, i.e., $\kappa$ is large. In this case, we can expect finite $\epsilon$ as per Eqn. 2. This is observed when a strong indentor force is applied by a metallic indentor in our experimental setup, extension exists in regions away from the indentor and sharp kinks in the region compressed by the indentor. The equilibrium configurations of the rods under various forces can be found by the balance of various forces on it. Forces on the element $ds$ are discussed in the main text.

## 3 Simulation

Initial Simulations were done in COMSOL using the Non-Linear Structural Materials module in COMSOL. However, this doesn't allow for frictional con-

tact as well as self-collisions were not possible to implement. The darker gray regions in the Fig. 5(b) in the main text therefore represent overlap of the material. Further finite element simulations were done in SOFA-Framework to take into account the self-collisions and friction coefficient $\mu$ was implemented at contacts. Codes can be found on [3].


\end{document}